\documentstyle[myepsfig]{aipproc}

\begin{document}

\title{Evidence for Centrifugal Barrier in X-ray Pulsar GRO~J1744-28}
 
\author{Wei Cui$^*$}
\address{$^*$Center for Space Research, Room 37-571, MIT, Cambridge, MA 02139}

\maketitle

\begin{abstract}
We present further observational evidence of the effects of a centrifugal 
barrier in GRO~J1744-28, based on continued monitoring of the source
with RXTE. 
\end{abstract}

\section*{Introduction}

The magnetosphere of an accreting X-ray pulsar expands as the mass accretion 
rate decreases. As it grows beyond the co-rotation radius, centrifugal force 
prevents material from entering it. Thus, accretion onto the magnetic poles 
ceases, and, consequently, X--ray pulsations cease. This phenomenon has 
recently been observed, for the first time, in GX~1+4 and GRO~J1744-28 with 
RXTE\cite{cui97}. Here, we present further evidence to show that the 
phenomenon repeated itself for GRO~J1744-28 during the decaying phase of its 
latest X-ray outburst.

\section*{Observations and Results}

The ASM light curve (as shown in the top panel of Fig.~1) reveals that there 
have been two episodes of X-ray outburst in GRO~J1744-28, separated by roughly
one year. The source has been extensively monitored by the main instruments 
aboard RXTE since its discovery\cite{fishman96}. For detailed analyses, we 
have selected a number of PCA observations, based on the ASM light curve, to 
cover the decay phase of the outbursts. Fig.~1 (bottom panel) shows the 
pulsed fraction ($\equiv (f_{max}-f_{min})/f_{max}$) measured with each 
observation. For comparison, the published results\cite{cui97} for the 
first outburst are also presented here. A striking feature is the precipitous 
drop of the pulsed fraction as the source became ``quiescent'' both times.

GRO~J1744-28 was generally not so quiet after the first outburst. In previous 
work\cite{cui97}, we happened to catch a brief period (as indicated in Fig.~1)
when the pulsed emission became very weak or was not detected at all in some 
observations. Following the latest ourburst, the 
source has shown little activity. Its presence (at about 20-30 mCrab) has, 
however, been firmly established by the PCA slew data. This provides a good 
opportunity to verify our previous interpretation of the phenomenon. We have 
searched for the known 2.14 Hz pulse frequency, employing various techniques 
including FFTs and epoch-folding, but have failed to detect it since the end 
of June 1997 (as marked in Fig.~1). The results therefore argue strongly that
the centrifugal barrier is active in this source during such faint period, as 
we have concluded previously\cite{cui97}. 
\begin{figure}[tb]
\epsfig{figure=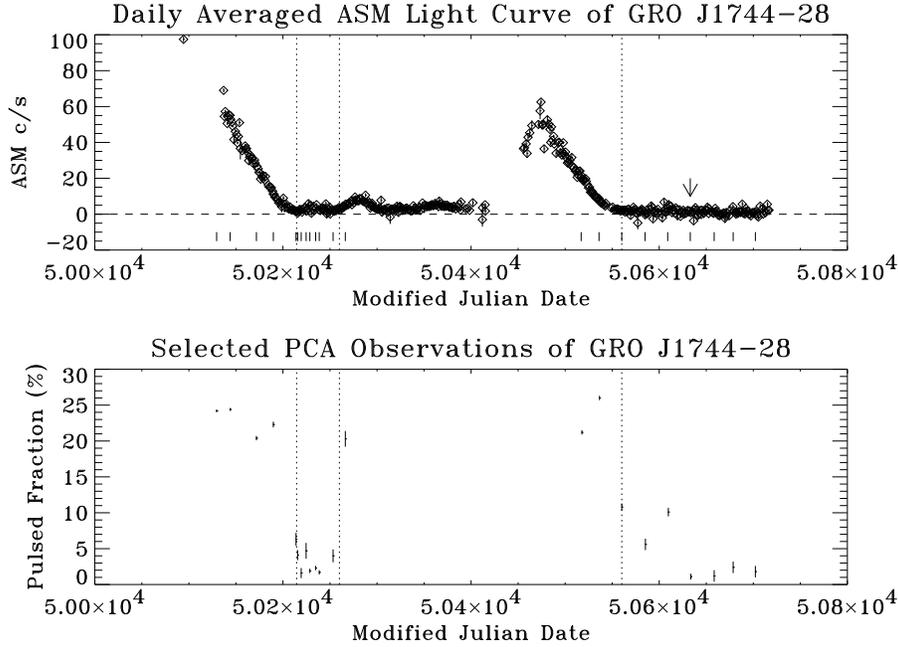,width=4.8in,angle=90}
\caption{(top) ASM light curve. The dotted lines indicate the start (or end) 
of the ``quiescent'' state, and the tick marks indicated the times when the 
selected PCA observations were made. The arrow marks roughly when the 
pulsation became undetectable following the latest outburst. (bottom) Measured
pulsed fraction (as defined in the text).}
\end{figure}

The source also shows interesting spectral evolution during the decay. The
observed X-ray spectrum can be characterized by a simple power law with an 
exponential high-energy cutoff. As the quiescent state is approached, the
spectrum softens significantly: the power-law becomes steeper, and more 
prominently, the cutoff energy decreases by roughly a factor of 2 (see 
Fig.~2). At the end of the first ``quiescent'' period, the spectrum would 
recover to the bright-state shape. We have proposed before that the X-ray 
emission probably 
consists of two components: the emission from a large portion of the neutron 
star surface (thus unpulsed), due to the ``leakage between field lines''
\cite{arons80}, and that from ``hot spots'' near the poles (pulsed plus 
unpulsed). When the source was bright, the latter dominated, so the spectrum 
was hard (corresponding to a much higher temperature of the hot spots). 
However, as soon as the centrifugal barrier took effect in the quiescent 
state, the observed X-rays were all due to the surface emission and their
spectrum was therefore softer.
\begin{figure}[tp]
\epsfig{figure=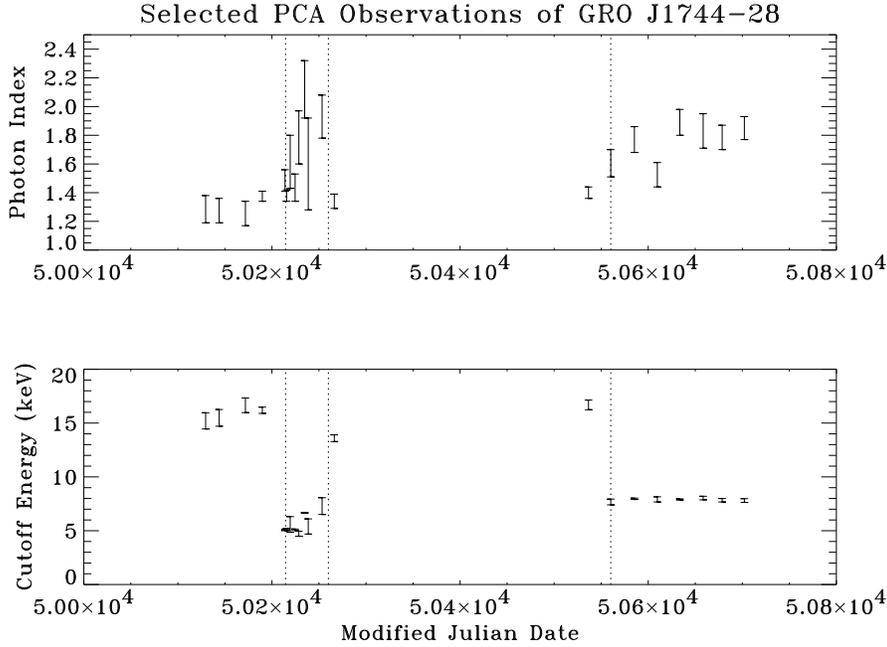,width=4.8in,angle=90}
\caption{Observed spectral evolution: (top) power-law photon index, and 
(bottom) high-energy cutoff.}
\end{figure}
\indent
It is interesting to note that the pileup of accreting matter on the neutron 
star surface might also cause unstable thermonuclear burning and produce 
type~I bursts\cite{bildsten97}, like in X-ray bursters. The lack of such (or 
does it?) in GRO~J1744-28 may be due to the suppression of this process by a 
significantly higher field\cite{rappaport96}. 

GRO~J1744-28 does produce X-ray bursts\cite{kou96a}, unlike any other X-ray 
pulsars. The bursts are thought to be the product of accretion 
instability\cite{lewin96}. They occurred at a rate of one to two dozen 
per hour near the peak of the outbursts\cite{kou96a,kou96b}, and the rate 
decreased as the X-ray flux decayed. At the start of the first quiescent 
period, the bursting activity ceased entirely\cite{kou96c} for weeks before 
resuming again near the end\cite{jahoda96}. Fig.~3 (the top panel) shows an 
example of such activity (with 7 major bursts) on MJD 50260 ($\equiv$ 26
June 1996).
\begin{figure}[tbp]
\epsfig{figure=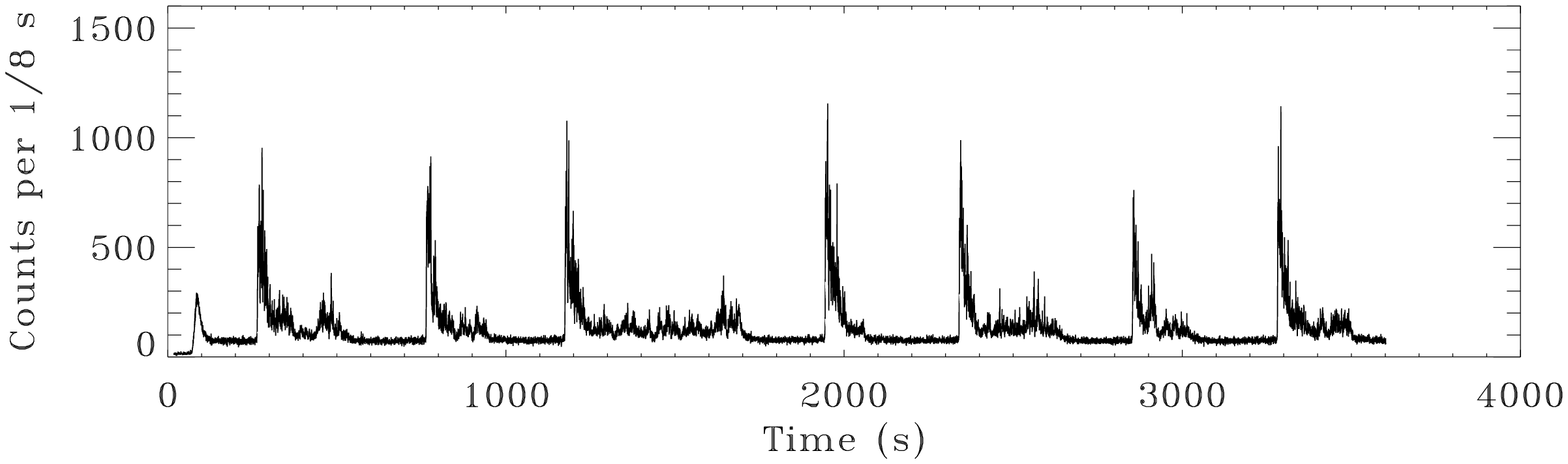,width=5in,angle=90}
\epsfig{figure=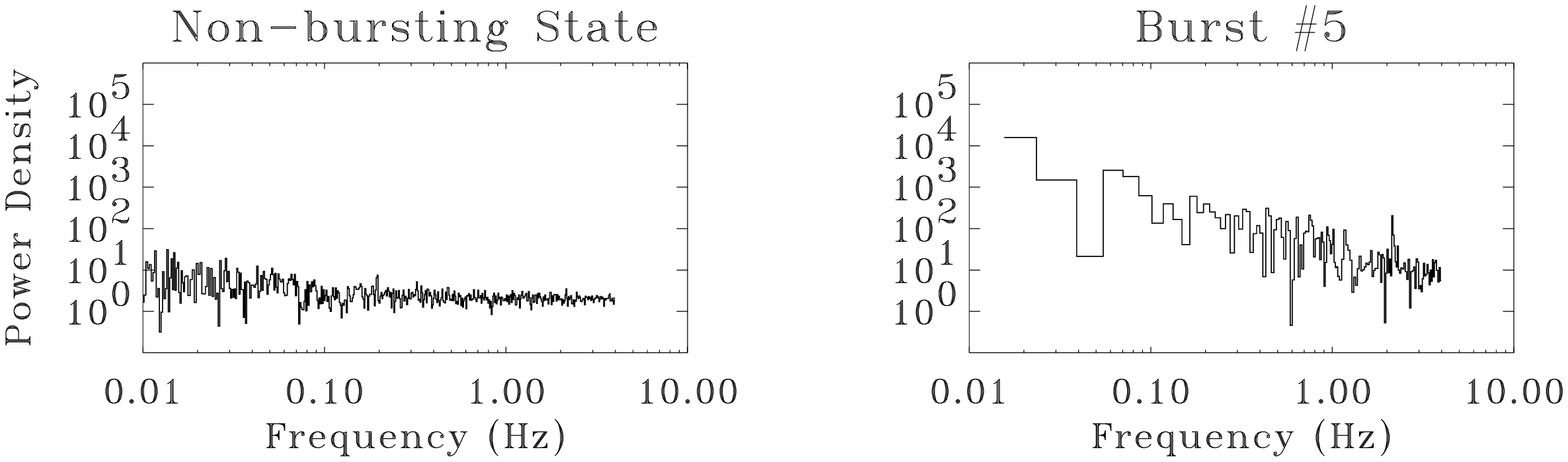,width=5in,angle=90}
\caption{(top) PCA light curve for the 26June1996 observation. (bottom) 
Power-density spectra for the non-bursting state and Burst \#5. Note the peak 
at 2.14 Hz in the latter.}
\end{figure}
We have separated the light curve of 26 June 1996 into burst and non-burst 
intervals. The X-ray pulsation is detected during the bursts but is {\it not} 
detected outside of them (see Fig.~3). This is again consistent with the 
presence of the centrifugal barrier 
in GRO~J1744-28. A sudden surge in the mass accretion rate that produces a 
burst would also momentarily push the magnetosphere inside the co-rotation 
radius and thus, the accretion to the poles would resume to produce the
pulsed emission. As the system relaxes following a burst, the magnetosphere 
expands again; the inhibition of accretion by the centrifugal barrier 
again suppresses the pulsation.

\section*{Conclusions}
\noindent
We conclude by summarizing the main results as follows:
\begin{itemize}
\item The results support our previous conclusion that the cessation of pulsed 
emission when the source becomes faint is a manifestation of the centrifugal 
barrier.
\item For GRO~J1744-28, the X-ray emission in the quiescent state (unpulsed) 
likely comes from 
a large portion of the neutron star surface, due to the penetration of 
accretion flows through the magnetosphere.
\item Accretion instability can still occur in the quiescent state (less
frequently), and produce type~II bursts. The pulsed emission was apparent 
during the bursts, presumably due to the resumption of accretion to the 
magnetic poles because of the momentary shrinkage of the magnetosphere.
The pulsation stopped as the system recovered to the quiescent state.
\end{itemize}

\end{document}